\documentclass[draft, 12pt, showpacs]{revtex4}
\usepackage{amssymb,amsmath}

\newcommand{\BR}{\mathbb{R}}
\newcommand{\SU}{\text{SU}}
\newcommand{\AdS}{\text{AdS}}
\newcommand{\YM}{\text{YM}}

\begin{document}
\title{Small AdS black holes from SYM}

\author{Curtis Asplund}
\email{casplund@physics.ucsb.edu}
\author{David Berenstein}
\email{dberens@physics.ucsb.edu}

 \affiliation{Department of
Physics, University of California
at Santa Barbara, CA 93106}

\begin{abstract}
We provide a characterization of the set of configurations in ${\cal N}=4$ SYM theory that
are dual to small AdS black holes.  Our construction shows that the black hole dual states are approximately thermal on a $\SU(M)$ subset of degrees of freedom of a $\text{SU}(N)$ gauge theory. $M$ is determined dynamically and the black hole degrees of freedom are dynamically insulated from the rest. These states are localized on the $S^5$ and have dynamical processes that correspond to matter absorption that make them behave as black objects.
\end{abstract}

\maketitle

The black hole information paradox has been a very interesting puzzle in theoretical physics \cite{Hawking}. 
The solution to this problem seems to require some amount of non-locality and that gravity should behave holographically \cite{'tHooft}. With the advent of the AdS/CFT correspondence \cite{Malda}, one would be inclined to state that the problem of unitarity in quantum gravity is resolved. To do this, one still needs a clean understanding of the dual configurations to black holes in field theory. If one finds such configurations, there is a lot that the field theory can teach us about the regime where semiclassical gravity breaks down, particularly with regards to the process of evaporating black hole dynamics. Large black holes are easily described by dual thermal states \cite{Witten}. 
The purpose of this paper is to give a set of configurations in field theory that correspond to the dual states of small black holes in AdS. Due to our approximations, we can only give very rough
features of these states and our results should be thought of as a caricature of the relevant dynamics of these systems.

Small Schwarzchild black holes in ten dimensions are allowed solutions of $\text{AdS}\times S$ gravity \cite{HH}. For very small black holes the local vicinity of the black hole is essentially flat space. These ten dimensional black holes are localized on the sphere geometry of $\AdS_5\times S^5$. They have negative specific heat and they evaporate. Yet, they survive for a very long time compared to their energy, suggesting that they are approximately metastable.  These black holes should be characterized by some dual set of states in the quantum field theory. These states must have some thermal behavior, be approximately metastable, highly entropic and yet correspond to a set of configurations that is far from equilibrium: the black holes evaporate after all. Thus, it is surprising that one can even assign a notion of temperature for them. In the field theory one needs a mechanism that isolates the degrees of freedom that make the black hole from the environment so that thermal physics makes sense. Also, these states are characterized essentially by a single number: the mass of the black hole in Planck units. 

We will argue using the ideas of emergent geometry described in \cite{BlargeN} that the configurations dual to a small black hole are given by  approximately thermal 
states in an $M\times M$  subsector of the large $N$ gauge field theory. The off-diagonal modes connecting this subsector to the rest of the field theory are heavy, providing some thermos that allows the $M\times M$ submatrix degrees of freedom to thermalize without spreading the thermal physics to all the $N\times N$ degrees of freedom. We will argue that $M$ and the temperature of the black hole are related. We will also show that these configurations localize in the $S^5$, that they are black and that they have negative specfic heat. This provides strong evidence for their identification as the dual states to small black holes in $\AdS$ space. 

\subsubsection{Semiclassical setup: setting the initial conditions}

In order to think of a black-hole in some spacetime, it is often useful to think of the black-hole as being formed by some dynamical process. If one can study the process of formation of a black hole, one might be able to account for the properties of the black hole itself and to understand what is the (complete) underlying description of the system. In situations like the AdS/CFT correspondence, this is a useful exercise, as the geometry of the ten dimensional spacetime is not immediately apparent in the field theory.

Our idea for studying black holes is to setup some initial conditions in the field theory that one can associate to configurations which would classically collapse and give rise to a black hole in the dual $\AdS$ setup. We then follow the configurations through collapse.

The idea for the initial conditions is simple: we study a dense gas of strings initially at rest in some small region
of $\AdS_5\times S^5$. This gas is centered on some location on the sphere, and at the bottom of $\AdS$ (we assume global $\AdS$ coordinates for this paper). This gas of strings is similar to non-relativistic dust in a small region, and we can choose the density of the gas to guarantee formation of a black hole once it collapses. We require that the associated black hole be very massive in string units, but also tiny with respect to the size of the sphere. These conditions require us to consider a large $N$ Super Yang Mills (SYM) at strong 't Hooft coupling. 

We now want to translate this initial condition to the field theory. To do this, we need a notion of geometry in field theory and also some notion of how to put many strings on it.
This dictionary was provided in \cite{BlargeN}. To find the geometry one begins by studying the ${\cal N}=4 $ SYM on $S^3\times \BR$. This is the boundary of global $\AdS$.
One then self-consistently decomposes the degrees of freedom into two classes: slow and fast, \`a la Born-Oppenheimer. The slow degrees of freedom correspond to a set of six commuting hermitian matrices, which are obtained from
just the s-wave modes of the scalar fields on the sphere \cite{BlargeN}. The fast degrees of freedom are everything else. Only the eigenvalues of the slow matrix degrees of freedom count as gauge independent. These eigenvalues can be organized as a collection of coordinates for $N$ particles in $\BR^6$, we label them $\mathbf{x}_i$.

There is a leading semiclassical effective
dynamics on these eigenvalues that is induced from the truncation to these degrees of freedom. The effective Hamiltonian is described by the following Schr\"odinger operator \cite{BlargeN}
\begin{equation}
H_{\text{eff}}= \sum_i -\frac 1{2\mu^2} \nabla_i \mu^2 \nabla_i + \frac 12 |\mathbf{x}_i|^2 ,
\end{equation}
where $\mu^2 = \prod_{i<j} |\mathbf{x}_i-\mathbf{x}_j|^2$. It was shown that for this system
the ground state wave function is given by 
\begin{equation}
\psi_0(\mathbf{x}_i) \sim \exp\left(-\frac12\sum |\mathbf{x}_i|^2\right)\label{eq:wf}.
\end{equation}
 In the large $N$ thermodynamic limit, the square of the wave function gives an associated probability for finding the particles in various locations on $\BR^6$. This can be best described by a density of eigenvalues. The measure gives an effective repulsion of the eigenvalues. The saddle point approximation gives a distribution of eigenvalues where all particles are located at distance
$r= \sqrt{N/2}$ from the origin \cite{BCV}, forming an $S^5$. This $S^5$ is interpreted as the 
sphere geometry that sits at the bottom of $\AdS$. In this setup the off-diagonal degrees of freedom are treated as if they are free on a first approximation (their frequencies depend on the slow degrees of freedom, but they don't interact with each other). Indeed, one uses this approximation for all other (fast) degrees of freedom. Each such mode is interpreted as a {\em string bit} connecting two eigenvalues. This string bit stretches between two points in the sphere (eigenvalues are interpreted as positions on the distribution of eigenvalues).  Due to gauge invariance one can show that the string bits are required to form closed polygons. This produces a collection of states that one can associated to closed strings.
One can reproduce exactly the energies of various strings in $\AdS_5\times S^5$ this way \cite{BCV}.

We would like to set up a gas of strings on some subregion of the sphere. We do this by 
choosing an initial condition where we stretch numerous string bits between various eigenvalues on a small spherical region of the $S^5$. We choose to use only s-wave degrees of freedom for simplicity, as these are all located at the bottom of AdS. Now we want to study the dynamical evolution of this system.

Our first problem is that the original description in terms of a wave function is too quantum mechanical to be of direct use. It would be better if one could treat the eigenvalue dynamics classically and think of them as moduli fields. The off-diagonal modes will be treated quantum mechanically. Since the geometry is intimately tied to 
what the eigenvalues are doing, classical eigenvalue dynamics is a natural proxy for the dynamics of the geometry. 

We perform a similarity transformation on $\psi$ by multiplying it by $\mu$. For the new wave function $\hat \psi=\mu\psi$, one obtains a different Hamiltonian, where the effective repulsion between the eigenvalues resides in the potential. This effective potential is given by
\begin{equation}	
	\label{Veff}
	V_{\text{eff}}	=\sum_i \frac 12\vert \mathbf{x}_i\vert^2  +\sum_{i, j\neq i}\frac{4}{\vert \mathbf{x}_{ij}\vert^2} + \frac{1}{2}\sum_i \left\vert \sum_{j\neq i} \frac{\hat{\mathbf{x}}_{ij}}{\vert \mathbf{x}_{ij}\vert} \right\vert^2,
\end{equation}
where $\mathbf{x}_{ij} = \mathbf{x}_i-\mathbf{x}_j$. It is easy to show that if one assumes a uniform distribution of particles, all of them at the same distance from the origin, then the minimum of the potential occurs at $r=\sqrt {N/2}$, up to $1/N$ corrections. Thus, the classical analysis of this effective potential already contains the quantum information that was stored in the wave function of equation \ref{eq:wf}. This accomplishes what we need: we can use classical physics to describe the eigenvalue dynamics. 

Turning on the string bits between the eigenvalues $i, j$ at occupation number $\hat N_{ij}$, gives an additional contribution to the Hamiltonian. With our free approximation this is given by
\begin{equation}
V_{\text{od}}= \sum_{ij} \hat N_{ij} \sqrt {1+ g_{\text{YM}}^2|\mathbf{x}_{ij}|^2} = \sum_{ij} \hat N_{ij} w_{ij},
\end{equation}
where $w_{ij}$ is the frequency of the corresponding oscillator. We assume that
the contribution that depends on $g_{\text{YM}}$ dominates the mass of the oscillator.
If the eigenvalues move slowly we can use the adiabatic approximation, taking $\dot N_{ij} \simeq 0$, and so treat the number of off-diagonal modes as constant.

 The interaction between the off-diagonal modes comes from commutator terms  in the $\mathcal{N}=4$ SYM potential. If one uses the off-diagonal harmonic oscillator raising and lowering operators,  one finds that the interactions  are suppressed by inverse powers of $w_{ij}$. So as long as the $w_{ij}$ are large, we can trust the approximately free description.
 
Our initial distribution has all particles at rest in a uniform distribution at radius $r=\sqrt{N/2}$, and a fixed collection of $\hat N_{ij}$ that affect all eigenvalues in a small region of the sphere.  We assume that $N_{ij}>0$ for a (possibly small) fraction of the pairs of eigenvalues in the chosen region. These string bits can stretch from one side of the region to the other.
 
 This is our dense gas of strings.
 We will assume that there are $M$ eigenvalues involved in the region of the sphere and that
 there is enough energy stored in this gas of strings so that in the dual AdS theory  it would collapse and form a large black hole in string units, but small compared to the size of the $S^5$. Thus $M \ll N$, but just to be sure, we want that $g_{\text{YM}}^2 M \gg 1$.
 
 The presence of $\hat N_{ij}$ takes the $M$ eigenvalues away from equilibrium: there is an extra force applied to them relative to the ground state. The energy of the off-diagonal modes decreases if we bring the eigenvalues closer to each other. Thus, the $M$ eigenvalues will tend to contract. This process converts potential energy stored in the string bits into kinetic energy for the eigenvalues and models collapse. It is mostly the static strings that want to shrink. For simplicity, we assume  
that there is no backreaction of the eigenvalues that are outside the region with strings. This backreaction should certainly be important if one wants to describe what happened to the geometry
of spacetime. However, at this stage we are just arguing that something special happens 
to this collapsing distribution: an object that is analogous to a black hole appears in the SYM theory. Notice also that for many other CFT's, the effective dynamics of eigenvalues in small regions is the same \cite{BHart}, so this process will be able to describe small black hole formation in any $AdS_5\times X$ geometry in a universal form.

\subsubsection{Thermalization and trapping}

Two things happen when we accelerate the eigenvalues towards each other. First, the masses of the off-diagonal modes change in time. If the eigenvalues are accelerated sufficiently towards each other we find that the adiabatic approximation breaks down. The parameter 
characterizing when this happens for each pair of eigenvalues is given by
\begin{equation}
\eta_{ij}= \frac{\dot w_{ij}}{w_{ij}^2}.
\end{equation}
When $\eta_{ij}>1$ there is particle production in the time dependent harmonic oscillators.
This will convert kinetic energy into the creation of extra string bits. This will slow down the acceleration of the eigenvalues and work as some effective friction. This pair creation also affects all  spherical harmonics of the field theory that have sufficiently fast time varying frequencies. 
On the other hand, the 
fact that more string bits are present make the forces pushing the eigenvalues together stronger.
Also notice that the
masses of string bits are becoming smaller, so the interactions between the string bits increase and the free string bit approximation breaks down.

We argue that this process will trap the eigenvalues into some soup-like state where the eigenvalues are much closer than the initial configuration.
This process will prevent the eigenvalues from scattering back to infinity in short times. This is analogous to 
the process of moduli-trapping in cosmology \cite{KLLMMS}. 

We further argue that the system can thermalize without affecting the outside eigenvalues
significantly. For, at the same time that the distances between the $M$ eigenvalues are shrinking, the distances to the outside eigenvalues are increasing. In particular, the 
lowest energies of string bits connecting the inside to the outside will be those connecting the inside eigenvalues to the edge of the 
region that was cut out by $M$.  The radius of the region is $R'$, and $(R'/R)^5 \sim M/N$. The frequency of these modes are of order 
\begin{equation}
w_{\text{in, out}}\sim g_{\YM} M^{1/5} N^{1/2-1/5} 
\end{equation}
as $N^{1/2}$ is the size of the sphere. This is a large number by our assumptions and it is much larger than the typical 
$w_{ij}$ for the eigenvalues that are going to be trapped. Thus, the production of string-bits 
between trapped and non-trapped particles is suppressed significantly. This guarantees that there is an effective thermos insulating the trapped degrees of freedom from the outside.

The dynamics of the trapped eigenvalues should become strongly interacting and chaotic. 
The coupling $g_{\YM}^2M$ being large and having large occupation numbers guarantees this. 
The presence of a field theory thermos guarantees that from a macroscopic point of view the notion of temperature for the trapped degrees of freedom makes sense. All we need is fast thermalization. 
It has been conjectured by Sekino and Susskind \cite{SS} that strongly interacting matrix models thermalize fast. We will assume that 
we can treat the $M$ trapped eigenvalues as a thermal system at a temperature $T$ after a short while. This is our caricature of a black hole. Also notice that once the moduli have rolled down, since the masses of string bits have been reduced significantly, we can distribute the energy between many more configurations. Thus,
the trapped state has a lot more entropy than the initial configuration and it will be hard for it to scatter back out. It will also be very non-abelian and as such can be considered as non-geometric. 

What is the effective thermodynamics? The $M$ eigenvalues slice a $\SU(M)$ subgroup of the original $\SU(N)$ SYM theory, corresponding to a block diagonal decomposition of matrices.  The $(N-M)\times (N-M)$ block is a  collection of diagonal matrices, and the $M\times M$ degrees of freedom are thermal. The $M\times M$ system should have the thermodynamics of 
a strongly coupled ${\cal N}=4 $ SYM theory. Thus, given $T,M$  we can estimate the entropy and the energy. We find this way that 
\begin{equation}
E \sim M^2 T^4, \ \ S\sim M^2 T^3.
\end{equation}
This is similar to the analysis \cite{BFKS} and is justified at strong coupling in \cite{GKT}.
Since this black hole is a thermal state of a subsector of the SYM theory, it is not particularly different from large black holes in AdS. It differs in that the system is not in thermal equilibrium with the environment and that it is localized on the sphere.

One can also expect that there will be Dirac-Born-Infeld (DBI) corrections to the eigenvalue dynamics \cite{Tseytlin}. These might be very important to describe the time scales of black hole formation and its properties and are part of the back-reaction problem.

We can also argue that the corresponding states behave as black objects. If we shoot matter towards the putative black hole, it will naturally carry some eigenvalue motion with it (this is how we can move in the geometry after all). 
If the outside matter gets sufficiently close to the black hole, then the off-diagonal modes connecting the black hole to the matter perturbation eigenvalues might be thermally accessible and might also suffer from non-adiabatic string bit pair creation. So long as the  typical trajectories of the matter perturbation stay near the black hole long enough, they are
trapped by the putative black-hole and will thermalize into it quickly. This is absorption by the black hole. Notice that DBI corrections in this case also have a thermal interpretation \cite{Silvertong}.

We can think of the special eigenvalues we have singled out, those that are `inside' the black hole, as D-branes. These D-branes are created out of the vacuum, and since the black hole is not charged they should be thought of as dielectric branes so that the black hole has no net number of branes. Counting black hole entropy by branes we find a picture that is reminiscent of other black-hole entropy counting arguments \cite{SV,BFKS}.

\subsubsection{Bounding the size and negative specific heat}

The last thing we need to determine is that the entropy and the energy should both depend on a single parameter (temperature, in this case), since this is a property of small black holes. So far, we have a two parameter description with $M$ and $T$ and these need to be related. 

The geometry of the $S^5$ sphere is constructed from typical field values. If we have a thermal state, we can  associate a size in the field directions by analyzing the thermal fluctuations of the field itself. This gives us a notion of the size of the black hole in the geometry. Standard thermal counting suggest that the thermal size of the black hole scales like $\delta \phi \sim \sqrt M T$. This size should be smaller than the region we cut out on the sphere. Otherwise the black hole grows and will eat more eigenvalues. Thus, we get an inequality of the form
\begin{equation}
\sqrt M T < (M/N)^{1/5} \sqrt N \label{eq:Tineq}.
\end{equation}
If we assume that the black hole grows to saturate some such inequality, then we can relate the temperature to $M$ and $N$: the number of trapped eigenvalues, and the Planck scale. The inequality \eqref{eq:Tineq} assumes no backreaction of the outside eigenvalues. There should be scaling corrections in $M/N$ to account for this and these should be very interesting to study. However, we can saturate \eqref{eq:Tineq} to get a very rough estimate.  
We find this way that $T\sim N^{3/10}M^{-3/10}$. Thus as we make the black hole smaller it heats up, as expected.
We also find that $S \sim N^{-3/4} E^{11/8}$. This is to be compared to the scaling for a Schwarzschild black hole in 10 dimensions: $S \sim E^{8/7}$. So our model has an exponent differing from this by about $20 \%$, which is not bad given the crudeness of our approximations.

\acknowledgements 
D.B. would like to thank G. Horowitz, D. Marolf, J. Polchinski and especially E. Silverstein for many discussions and correspondence. Work supported in part by DOE under grant DE-FG02-91ER40618.


\begin{thebibliography}{99}

\bibitem{Hawking}
  S.~W.~Hawking,
  Commun.\ Math.\ Phys.\  {\bf 43}, 199 (1975)
  [Erratum-ibid.\  {\bf 46}, 206 (1976)].
  Phys.\ Rev.\  D {\bf 14}, 2460 (1976).
  
\bibitem{'tHooft}
  G.~'t Hooft,
  arXiv:gr-qc/9310026. Salamfest 1993:0284-296
  L.~Susskind,
  J.\ Math.\ Phys.\  {\bf 36}, 6377 (1995)
  [arXiv:hep-th/9409089].

\bibitem{Malda}
  J.~M.~Maldacena,
  Adv.\ Theor.\ Math.\ Phys.\  {\bf 2}, 231 (1998)
  [Int.\ J.\ Theor.\ Phys.\  {\bf 38}, 1113 (1999)]
  [arXiv:hep-th/9711200].

\bibitem{Witten}
  E.~Witten,
  Adv.\ Theor.\ Math.\ Phys.\  {\bf 2}, 253 (1998)
  [arXiv:hep-th/9802150].
  Adv.\ Theor.\ Math.\ Phys.\  {\bf 2}, 505 (1998)
  [arXiv:hep-th/9803131].

\bibitem{HH}
  G.~T.~Horowitz and V.~E.~Hubeny,
  JHEP {\bf 0006}, 031 (2000)
  [arXiv:hep-th/0005288].


\bibitem{BlargeN}
  D.~Berenstein,
  JHEP {\bf 0601}, 125 (2006)
  [arXiv:hep-th/0507203].

\bibitem{BCV}
  D.~Berenstein, D.~H.~Correa and S.~E.~Vazquez,
  JHEP {\bf 0602}, 048 (2006)
  [arXiv:hep-th/0509015].

\bibitem{BHart}
  D.~Berenstein,
  JHEP {\bf 0804}, 002 (2008)
  [arXiv:0710.2086 [hep-th]].
  D.~E.~Berenstein and S.~A.~Hartnoll,
  JHEP {\bf 0803}, 072 (2008)
  [arXiv:0711.3026 [hep-th]].
  
  
\bibitem{Tseytlin}
  I.~L.~Buchbinder, A.~Y.~Petrov and A.~A.~Tseytlin,
  Nucl.\ Phys.\  B {\bf 621}, 179 (2002)
  [arXiv:hep-th/0110173] and references therein.
  
\bibitem{KLLMMS}
  L.~Kofman, A.~Linde, X.~Liu, A.~Maloney, L.~McAllister and E.~Silverstein,
  JHEP {\bf 0405}, 030 (2004)
  [arXiv:hep-th/0403001].
  
\bibitem{SS}
  Y.~Sekino and L.~Susskind,
  arXiv:0808.2096 [hep-th].

\bibitem{Silvertong}
  M.~Alishahiha, A.~Karch, E.~Silverstein and D.~Tong,
  AIP Conf.\ Proc.\  {\bf 743}, 393 (2005)
  [arXiv:hep-th/0407125].

\bibitem{BFKS}
  T.~Banks, W.~Fischler, I.~R.~Klebanov and L.~Susskind,
  Phys.\ Rev.\ Lett.\  {\bf 80}, 226 (1998)
  [arXiv:hep-th/9709091].
  
\bibitem{GKT}
  S.~S.~Gubser, I.~R.~Klebanov and A.~A.~Tseytlin,
  Nucl.\ Phys.\  B {\bf 534}, 202 (1998)
  [arXiv:hep-th/9805156].
  
\bibitem{SV}
  A.~Strominger and C.~Vafa,
  Phys.\ Lett.\  B {\bf 379}, 99 (1996)
  [arXiv:hep-th/9601029].
  U.~H.~Danielsson, A.~Guijosa and M.~Kruczenski,
  JHEP {\bf 0109}, 011 (2001)
  [arXiv:hep-th/0106201].

\end{thebibliography}
\end{document}